\documentclass[aps,prb,preprint,superscriptaddress]{revtex4-1}

\usepackage{epsfig,verbatim,hyperref,color,float}
\usepackage{amsmath,amsfonts,mathrsfs,natmove}


\usepackage{xcolor}

\begin{document}

%\preprint{AIP/123-QED}

\title{The universal emergence of self-affine roughness from deformation}
\vspace{2cm}
\author{{ Adam R. Hinkle$^{1,2,*}$, Wolfram G. N\"ohring$^{3,*}$ and Lars Pastewka$^{2,3,4}$}\\
{\small \em 
$^1$Materials, Physical and Chemical Sciences Center, Sandia National Laboratories, Albuquerque, NM 87123, USA\\
$^2$Institute for Applied Materials, Karlsruhe Institute of Technology, 76131 Karlsruhe, Germany \\
$^3$Department of Microsystems Engineering, University of Freiburg, 79110 Freiburg, Germany\\
$^4$Freiburg Materials Research Center, University of Freiburg, 79104 Freiburg, Germany\\
$^*$These authors contributed equally.
}}

\begin{abstract}
Most natural and man-made surfaces appear to be rough on many length scales. There is presently no unifying theory of the origin of roughness or the self-affine nature of surface topography. One likely contributor to the formation of roughness is deformation, which underlies many processes that shape surfaces such as machining, fracture, and wear. Using molecular dynamics, we simulate the bi-axial compression of single-crystal Au, the high-entropy alloy Ni$_{36.67}$Co$_{30}$Fe$_{16.67}$Ti$_{16.67}$, and amorphous Cu$_{50}$Zr$_{50}$, and show that even surfaces of homogeneous materials develop a self-affine structure. By characterizing subsurface deformation, we connect the self-affinity of the surface to the spatial correlation of deformation events occurring within the bulk and present scaling relations for the evolution of roughness with strain.
\end{abstract}

\maketitle

Surface roughness~\cite{persson_roughness} appears across many length scales and in almost all physical systems, including the rocky terrain of mountain ranges~\cite{Gagnon2006-sb}, metals~\cite{mandelbrotnature,zaiser,wouters}, glasses,~\cite{Ponson2006-gn} and silicon wafers~\cite{khan}. Roughness critically controls friction~\cite{Urbakh2004-pb}, adhesion~\cite{Pastewka2014}, and transport~\cite{Binder1912,Gotsmann2013}, and plays a decisive role in both industrial and scientific fields, from operating machinery to predicting earthquakes. Rough surfaces are often fractals with statistical self-affine scaling~\cite{mandelbrot,Sayles1978-ja} observed from the atomic to the tectonic~\cite{Gagnon2006-sb,ninedec,jacobsuncd}. There is presently no unifying explanation for the origins of this self-affinity, but the influence of microstructural heterogeneity on material deformation is widely cited as a possible mechanism.~\cite{Becker1998-wa,Zhao2004-lw,Yue2005-ja,wouters,Sundaram2012-lm,Sandfeld2014-he}

The fact that scale-invariant roughness is observed from microscopic to geological scales hints that a common mechanism is active across vastly different length scales. This is surprising since the processes that form mountain ranges or the surface of a ball bearing are excruciatingly complicated. Geological faults crack, slide, and wear, and man-made surfaces typically undergo many steps of shaping and finishing, such as polishing, lapping, and grinding. Yet, all of these surface changes, whether natural or engineered, involve mechanical deformation at the smallest scales: Even the crack-tips of most brittle materials such as glasses exhibit a finite process-zone where the material is plastically deformed.~\cite{Alava2006-nu} This smallest scale of roughness is important because it controls the contact area~\cite{persson_roughness} and thereby adhesion~\cite{Pastewka2014}, conductance~\cite{Gotsmann2013}, and other functional properties.

\begin{figure}
\hspace{-0.5cm}
\includegraphics{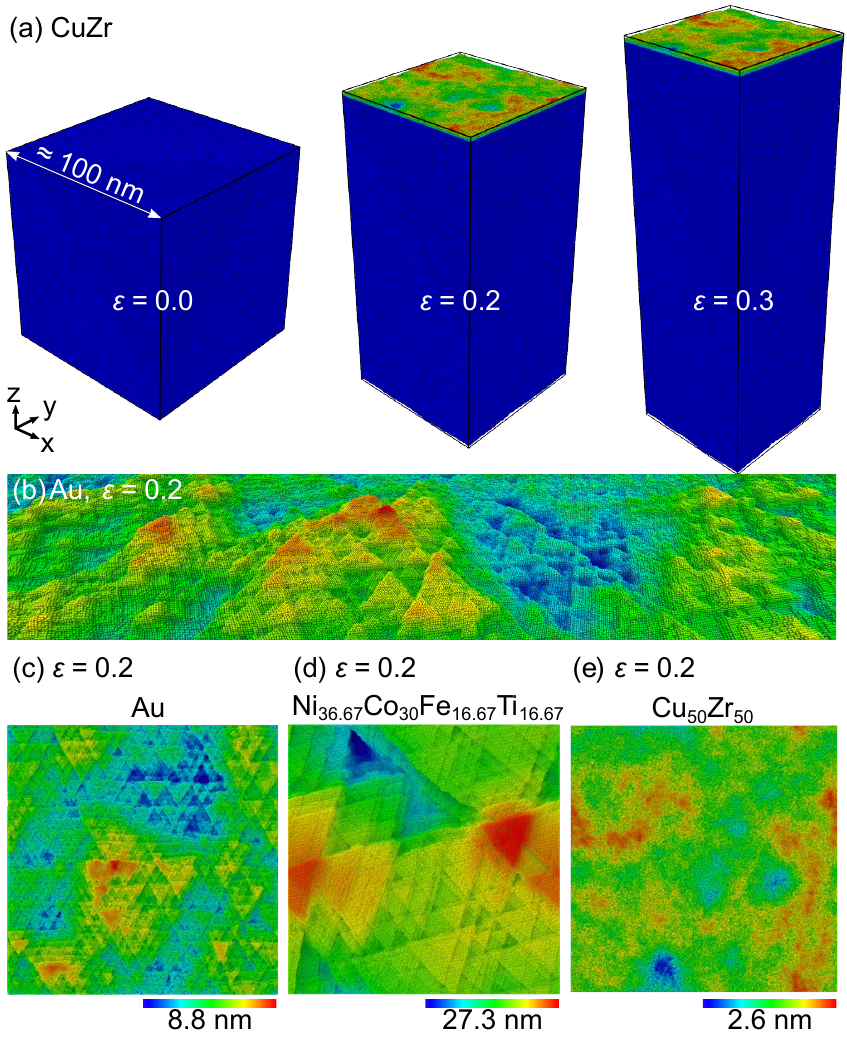}
\caption{Molecular dynamics simulations of the formation of surface roughness in single-crystal Au and the high-entropy alloy Ni$_{36.67}$Co$_{30}$Fe$_{16.67}$Ti$_{16.67}$, both with a (111) surface orientation, and amorphous Cu$_{50}$Zr$_{50}$. (a) Evolution of the full simulation cell during compression of amorphous CuZr, illustrating the simulation protocol. During compression the surface of initial area $100~ \mu\text{m}\times 100~\mu\text{m}$ roughens. The color encodes the atomic position normal to the surface measured relative to the surface's mean height. (b) Perspective view of the surface roughness that develops on Au. Bottom row shows topography maps of (c) Au, (d) NiCoFeTi, and (e) CuZr. Panels (b-e) are at an applied strain of $\varepsilon=0.2$. Panel (b) and (c) share the same color map.}
\label{fig:surfaces}
\end{figure}

In this Letter, we report molecular dynamics (MD) calculations of simple biaxial compression for three benchmark material systems: single-crystal Au, the model	 high-entropy alloy Ni$_{36.67}$Co$_{30}$Fe$_{16.67}$Ti$_{16.67}$, and amorphous Cu$_{50}$Zr$_{50}$. Each system is known to exhibit a different micromechanical or molecular mechanism of deformation, and we study the ensuing atomic-scale changes both within the bulk of the system and the emerging rough surfaces during the applied compression (see Fig.~\ref{fig:surfaces}a and Methods). The systems initially consist of cubic volume elements with lateral length $L\approx 100$~nm, and each material represents a unique limit of structural order: a homogeneous crystal, a crystal with stoichiometric disorder, and a glass with no long-range order. Despite their differences in structure and material properties, all three systems develop rough surfaces when biaxially compressed as shown in Fig.~\ref{fig:surfaces}.

\begin{figure}
\begin{center}
\includegraphics{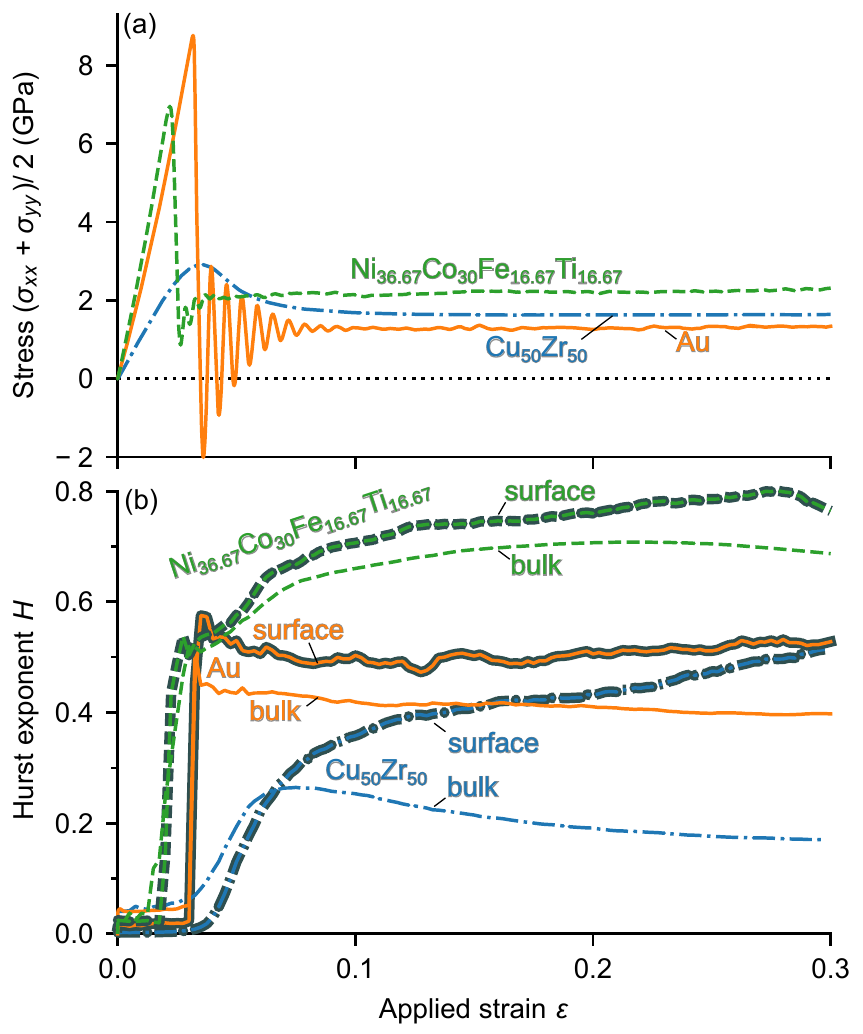}
\caption{Analysis of the deformation process as a function of applied strain $\varepsilon$. (a) Stress during deformation. (b) Hurst exponent $H$ at the surface and within the bulk computed from a fit to the scaling analysis.}
\label{fig:stress}  
\end{center}
\end{figure}

The stress-strain response of our systems during this process is typical (Fig.~\ref{fig:stress}a): Stress increases linearly in the elastic regime until yielding begins. Because our crystalline systems are homogeneous on scales beyond a few atomic distances and contain no pre-existing defects, the yield stress is much larger than the stress at which they flow~\cite{Zepeda-Ruiz2017-hx}. The amorphous system shear-softens as is typical for metallic glasses.~\cite{Hufnagel2016-bb}

\begin{figure}
\begin{center}
\includegraphics[width=15cm]{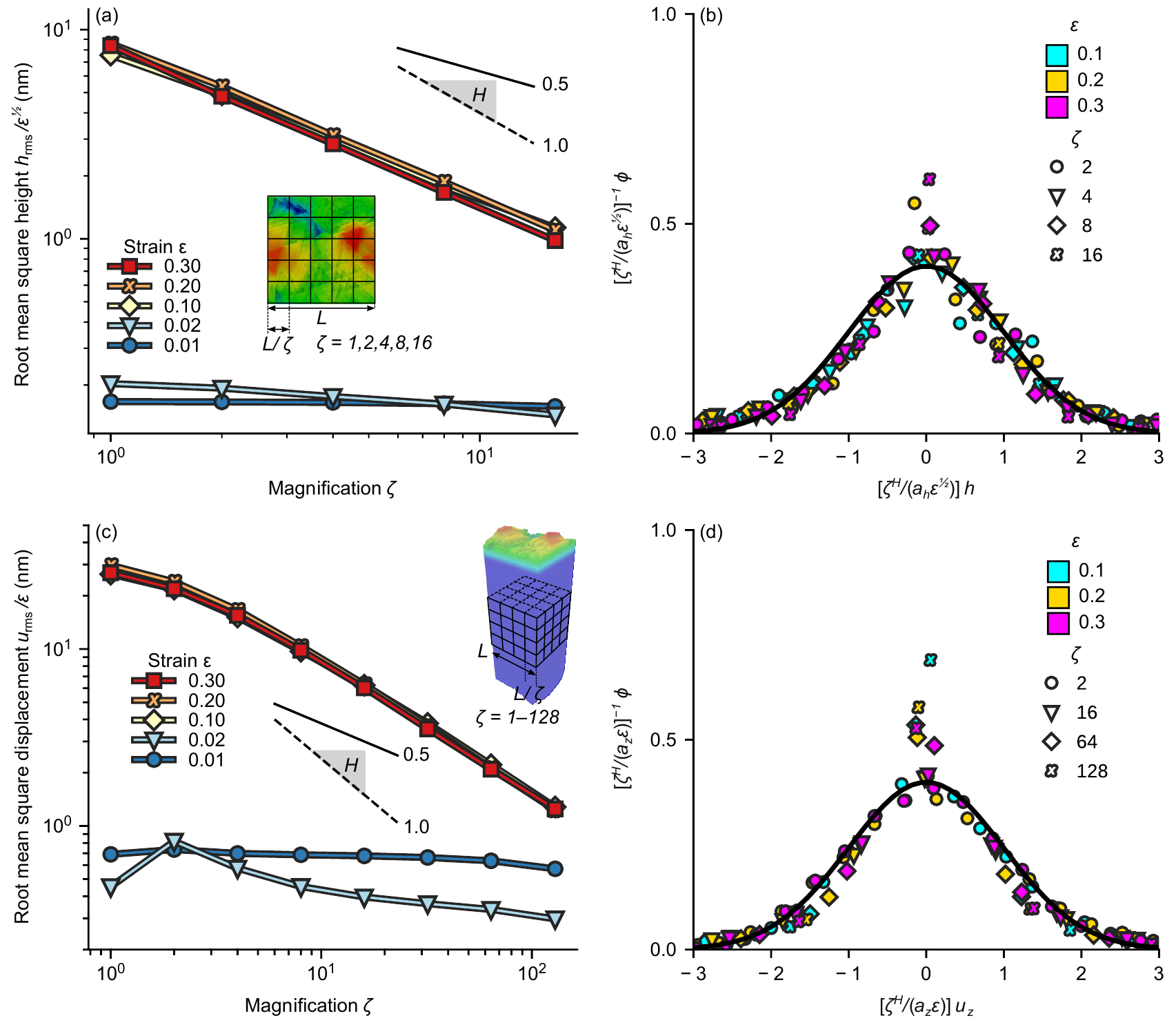}
\caption{Detailed analysis of the surface topography of NiCoFeTi. (a) Root-mean-square height $h_\text{rms}$ as a function of magnification $\zeta$ showing self-affine scaling over more than one decade in length. The data collapse in the plastic regime when normalized by $\varepsilon^{1/2}$, where $\varepsilon$ is the strain due to compression. Panel (b) shows the underlying distribution function $\phi_\zeta(h;\varepsilon)$ at different $\varepsilon$, which collapses upon rescaling heights $h$ by $\zeta^H/(a_h \varepsilon^{1/2})$ and letting $a_h=6$~nm. (c) Root-mean-square amplitude $u_\text{rms}$ of the $z$-component of the subsurface displacement field $u_z$ as a function of $\zeta$ within the bulk. The displacement data collapse when normalized by $\varepsilon$. The bulk displacement field shows self-affine scaling over more than two decades in magnification. Panel (d) shows the underlying distribution function of the displacements $u_z$, which collapses upon rescaling displacements $u_z$ by $\zeta^H/(a_z \varepsilon)$ and letting $a_z=34$~nm.  Solid and dashed lines in (a) and (c) show perfect self-affine scaling for reference with $H=0.5$ and $H=1.0$, respectively. The solid lines in panels (b) and (d) show the standard normal distribution. The respective scaling collapses use the same Hurst exponent $H=0.7$.}
\label{fig:psd_cuzr_au}  
\end{center}
\end{figure}

Ideal self-affine roughness is characterized by statistical scale-invariance.~\cite{mandelbrot} Following a classical procedure,\cite{Mandelbrot1985-af} we subdivide our surfaces into checkerboard patterns of squares with length $L/\zeta$ and compute the height distribution functions $\phi_\zeta(h;\varepsilon)$ for different magnifications $\zeta$ as the mean over all squares (see Methods). The surface is statistically self-affine, if the height distribution at magnification $\zeta$ corresponds to the one at $\zeta=1$ but with all heights rescaled by $\zeta^{-H}$, where $H$ is the Hurst exponent.~\cite{persson_roughness} Figure~\ref{fig:psd_cuzr_au}a shows the root mean square height $h_\text{rms}$ (the standard deviation of $\phi_{\zeta}$) namely $h_{\text{rms},\zeta}(\varepsilon)=\left\{\int dh\;h^2 \phi_\zeta(h;\varepsilon)\right\}^{1/2}$, at particular magnifications. It scales as $h_{\text{rms},\zeta} \propto \zeta^{-H}$ and the data sets for different $\varepsilon$ collapse if $h_{\text{rms}, \zeta}$ is normalized by $\varepsilon^{1/2}$. The self-affine scaling and collapse with $\varepsilon^{1/2}$ also holds for Au and CuZr (Supplementary Material, Figs.~S-1 and S-2). We believe that the scaling $h_{\text{rms},\zeta}(\varepsilon) \propto \varepsilon^{1/2}$ is the signature of an emerging surface roughness that is uncorrelated in strain. A detailed discussion on the underlying atomic-scale mechanisms and arguments for this scaling behavior can found in Supplementary Section S-I.

Applying this analysis to individual snapshots during the deformation allows us to follow the evolution of $H$ with $\varepsilon$ (Fig.~\ref{fig:stress}b). In the elastic regime, the surfaces are not self-affine. This is manifested in an $h_{\text{rms},\zeta}$ that is independent of magnification $\zeta$, leading to a Hurst exponent $H=0$ (Fig.~\ref{fig:psd_cuzr_au}a). The Hurst exponent jumps to a value around $H\sim 0.5$ for Au and NiCoFeTi at yield. A value of $H=0.5$ indicates a random walk, i.e. uncorrelated slip lines from dislocations that annihilate at the surface. Upon further deformation of NiCoFeTi, $H$ evolves to values $0.5<H<0.8$ indicating that the nucleation and motion of dislocations becomes increasingly correlated for this material. For the amorphous system, $H$ smoothly evolves from a value at yield $H=0.4$ to $0.5$ at $30\%$ strain. The Hurst exponent of the amorphous system is strongly temperature dependent (Supplementary Fig.~S-3) while the results for the crystalline systems are robust over a range of temperatures. We note that similar values for the Hurst exponent have been reported for stochastic crystal plasticity models~\cite{Zaiser2005-vm} and observed in compression experiments carried out on polycrystalline Cu,~\cite{zaiser} cleaved optical-grade KCl,~\cite{schwerdt} and LiF~\cite{Schwerdtfeger2010-zm}. No similar experimental data presently exist for high-entropy alloys or amorphous systems.

These results show that $H$ varies only weakly as the material flows. This encourages us to attempt a collapse of the full height distribution
\begin{equation}
\phi_\zeta(h;\varepsilon)=\frac{\zeta^{H}}{a_h \varepsilon^{1/2}} f\left(\frac{\zeta^{H}}{a_h \varepsilon^{1/2}} h\right)
\end{equation}
for different $\zeta$ and $\varepsilon$ onto a universal scaling function $f(x)$ with a constant Hurst exponent $H$ and length-scale $a_h$. Figure~\ref{fig:psd_cuzr_au}b shows this collapse for NiCoFeTi with $H=0.7$ and $a_h=6$~nm. The underlying scaling function $f(x)$ can be approximated by the standard normal distribution. The deformation of Au (Fig.~S-1) and CuZr (Fig.~S-2) shows identical behavior, albeit with different $H$ and $a_h$.

We further recognize that the surface topography $h(x,y)$ is simply the normal component of the displacement field $\vec{u}(x,y,z)$ evaluated at the surface, $h(x,y)\equiv u_z(x,y,z=0)$. This raises the question as to whether the self-affine structure is a signature of the deformation within the bulk itself. To address this question we carry out identical scaling analyses on the bulk displacement field in the center of the simulation box, away from the surface (see Methods). In this now three-dimensional ``topography'', the magnification $\zeta$ refers to a cubic discretization of space (inset to Fig.~\ref{fig:psd_cuzr_au}c). For NiCoFeTi (Fig.~\ref{fig:psd_cuzr_au}c), Au (Fig.~S-1c), and CuZr (Fig.~S-2c), and we indeed find a self-affine displacement field. Similar to the topography analyses, the full scale-dependent distributions of the $z$-component of the displacement can be collapsed onto a single scaling function, which can 
be approximated by the standard normal distribution (Figs.~\ref{fig:psd_cuzr_au}d, S-3d and S-2d). Unlike the topography, the root mean square fluctuation of $u_z$ scales as $\varepsilon$ (see Supplementary Section S-I for a discussion). The Hurst exponents extracted from the subsurface and the surface are identical for NiCoFeTi and Au (Fig.~\ref{fig:stress}c). For CuZr, we find a smaller Hurst exponent in the subsurface region which we attribute to self-diffusion within the glass that is driven by the applied strain~\cite{Berthier2002-eg} but absent in crystals. Bulk and surface diffusion is likely also the reason for the strong temperature dependence of $H$ in the amorphous system. Further work is necessary to quantitatively describe the influence of diffusion on the fractal nature of the topography and displacement field.

We note that the occurrence of a self-affine geometry in the displacement field is compatible with previous observations regarding the spatial correlations of noise sources during the creep deformation of ice.~\cite{weiss} Deformation does not manifest as smooth laminar flow, and the statistical nature of plasticity~\cite{Zaiser2006-rs} is the key reason why surfaces develop self-affine roughness during deformation. It is remarkable that deformation in crystalline solids shows statistical scaling identical to amorphous solids, despite the fact that plasticity is carried by shear transformations~\cite{Argon,Spaepen,falk2} in amorphous, and by dislocations~\cite{Zaiser2005-vm,Zaiser2006-rs} in crystals, two topologically distinct defects. This suggests that the emergence of self-affine roughness is independent of the deformation mechanism and carries over to deformation processes occurring at much larger scales, such as those in geology. Values in the range of $0.5<H<0.8$ are found on fracture surfaces~\cite{mandelbrotnature,Ponson2006-gn}, mountain ranges~\cite{Gagnon2006-sb}, geological faults~\cite{ninedec} and deformed crystals~\cite{zaiser,schwerdt,Schwerdtfeger2010-zm,wouters} and we find values in this range for molecular dynamics simulations of deformed crystals and bulk metallic glasses at low temperature. Since all our calculations are carried out on homogeneous systems without internal regions over which homogeneity is broken, such as grains or precipitates, our calculations clearly demonstrate that material heterogeneity is not a necessary prerequisite for the emergence of self-affine roughness. While heterogeneity, such as crystalline grains, does affect how materials accomodate deformation~\cite{Becker1998-wa,Zhao2004-lw,Yue2005-ja,Sundaram2012-lm}, our results explain why self-affine roughness is found to extend to subgrain scales.~\cite{wouters}

The results of this work shed light on the origin of self-affine surface roughness and its connection to deformation by systematically studying atomistic calculations of homogeneous solids with varying degrees of disorder. Our approach, using molecular dynamics to probe the evolving material surfaces, allows examination of the evolution of the Hurst exponent throughout the entire process of deformation, not only at free surfaces but anywhere within the material. In particular, we present quantitative evidence that self-affine surface roughening is linked to the statistical mechanics of deformation and derive scaling expressions that describe the evolution of self-affine roughness with strain in terms of just two parameters, the Hurst exponent and an internal length scale. Our results pave the way for a thorough understanding and control of surface roughness created in a variety of processes, such as machining or wear.

\emph{Acknowledgements.} We thank Tevis Jacobs, Laurent Ponson, Daniel Weygand and Michael Zaiser for useful discussions. The authors acknowledge support from the Deutsche Forschungsgemeinschaft (Grant PA 2023/2) and the European Research Council (ERC-StG-757343). We are indebted to the J\"{u}lich Supercomputing Center for allocation of computing time on JUQUEEN and JUWELS (grant hka18). Post-processing was carried out on NEMO at the University of Freiburg (DFG grant INST 39/
963-1 FUGG). All simulations were carried out with LAMMPS~\cite{lammps} and post-processed with OVITO~\cite{ovito}.

\emph{Methods.} The initial configuration of crystalline Au is an ideal fcc slab. The high-entropy alloy consists of random elements distributed on an fcc lattice. Both crystalline systems have a $(111)$ surface orientation. The Au system is oriented along the $x$- and $y$-axes in $[\bar{1}10]$ and $[\bar{1}\bar{1}2]$ directions, respectively. Preliminary calculations on the high-entropy alloy using the same lattice orientation showed the formation of a shear band parallel to the periodic simulation cell faces. To suppress this shear-band, we rotate the lattice to $[\bar{3}4\bar{1}]$ and $[\bar{5}\bar{2}7]$ directions in the $x$- and $y$-axes. The CuZr glass is formed by taking a 50-50 composition of the binary alloy and quenching the liquid equilibrated for $100$~ps at a temperature of $1800$~K at a rate of $10^{11}$~K~s$^{-1}$. The systems have a linear dimension $L\approx 100$~nm.  The CuZr system contains $58$ million atoms, the Au system $60$ million, and the NiCoFeTi system $83$ million. 
Atoms interact via embedded atom method (EAM) potentials in all three cases: Grochola et al.~\cite{grochola} for Au, Zhou et al.~\cite{Zhou2001-mc} for NiCoFeTi, and Cheng et al.~\cite{eam} for CuZr. The potential by Zhou et al. \cite{Zhou2001-mc} was recently used by Rao et al. \cite{rao_atomistic_2017} to study glide of single edge and screw dislocations in Ni$_{36.67}$Co$_{30}$Fe$_{16.67}$Ti$_{16.67}$ and should be regarded as a model of a complex solid solution alloy with a high concentration of the individual components and a stable fcc phase. 
Both the amorphous and crystalline systems are subjected to the same biaxial compression protocol: We apply a constant strain rate $\dot{\varepsilon}_{xx}=\dot{\varepsilon}_{yy}= -10^{8}$~s$^{-1}$ by uniformly shrinking dimensions of the simulation box along the $x$ and $y$ directions. To eliminate artifacts during compression that can occur for large systems in molecular dynamics, we ramp the strain rate smoothly to the final rate over a time interval of $100$~ps and employ a momentum conserving thermostat (Dissipative Particle Dynamics, e.g. Ref.~\onlinecite{Soddemann2003-el}) with a relaxation time constant of roughly $1$~ps. Unless otherwise noted, simulations are carried out at a temperature of $100$~K.

To extract the profile of the rough surface $h(x,y)$, we subdivide the surface into quadratic bins of linear size $d$. The height $h$ within each bin is the $z$-position of the top-most atom. We systematically check the influence of $d$ which must be larger than the nearest-neighbor spacing between atoms. All calculations are analyzed with $d=3$~\AA. For the subsequent scaling analysis, we subdivide $h(x,y)$ into regular square cells of size $L/\zeta$ (inset to Fig.~\ref{fig:psd_cuzr_au}a) and tilt-correct through affine deformation the rough profile within each cell individually before computing the full height distribution function and the rms height within each cell. The final distribution function $\phi_\zeta(h)$ and rms height $h_\text{rms}(L/\zeta)$ is computed as the mean over all cells.

The non-affine part of the displacement in the bulk for each atom $i$ at strain $\varepsilon$ is obtained as $\vec{u}_i(\varepsilon)=\vec{r}_i(\varepsilon)-\underline{F}(\varepsilon)\vec{r}_i(0)$, where $\vec{r}_i(\varepsilon)$ is the position of atom $i$ at applied strain $\varepsilon$. The tensor $\underline{F}(\varepsilon)$ is the deformation gradient that transforms the initial system at applied strain $\varepsilon=0$ to the current state. Analysis of $u_{z,i}$, the $z$-component of $\vec{u}_i$, is carried out along the lines of the roughness analysis. We subdivide a cube centered in the middle of the deformed system into cubes of size $L/\zeta$ (inset to Fig.~\ref{fig:psd_cuzr_au}c) and compute the distribution $\phi_\zeta(u_z)$ and rms fluctuation after removing the affine part of the deformation within each cube individually (the ``tilt'' correction of the displacement field). Removal of the affine part of the deformation field is carried out as in Ref.~\onlinecite{falk2} but within cubes and not augmentation spheres around atoms. The final $\phi_\zeta(u_z)$ and $u_\text{rms}(L/\zeta)$ is the mean of the individual quantities for each of the cubes.

%\bibliography{Hinkle_Noehring_Pastewka}

\newpage

\setcounter{figure}{0}
\renewcommand{\thesection}{S-\Roman{section}}
\renewcommand{\thefigure}{S-\arabic{figure}}
\renewcommand{\thetable}{S-\arabic{table}}
\renewcommand{\theequation}{S-\arabic{equation}}

\begin{center}
\huge\bf{ Supplementary Material for \\
 ``The universal emergence of self-affine roughness from deformation'' }

\vspace{1cm}

\large Adam R. Hinkle$^{1,2}$, Wolfram G. N\"ohring$^{3}$ and Lars Pastewka$^{2,3}$
\end{center}

$^1$Sandia National Laboratories, Albuquerque, NM 87185, USA

$^2$Karlsruhe Institute of Technology, 76131 Karlsruhe, Germany

$^3$University of Freiburg, 79110 Freiburg, Germany

\section{Atomic-scale deformation mechanisms}

For the sake of illustration, we here discuss the crystalline systems. In our crystalline fcc systems, deformation occurs by slip on $(111)$ planes. There are three $(111)$ planes, all oriented at the tetrahedral angle ($\alpha_t=109.47°$) with respect to each other. A full dislocation that annihilates at the surface leaves behind a step of height $\Delta_\perp = a_0/\sqrt{3}$ where $a_0$ is the lattice constant of the crystal. During compression, the crystal will shrink by a distance $\Delta_\parallel = \Delta/\tan(\pi-\alpha_t) = a_0/\sqrt{24}$ for each surface step. Given a linear dimension $L$, compression by $\varepsilon$ will hence give rise to $N=L\varepsilon/\Delta_\parallel$ steps on the surface.

We now regard two limits of this process: In limit A, these steps occur at random positions on the surface. The surface profile then constitutes a random walk (and the Hurst exponent would be 0.5). Because the surface remains nominally flat, the walk needs to be self-returning. This happens either due to lattice rotation or because the stress introduced at the surface when creating a single step makes it more likely that the next step is in the opposite direction.~\cite{Sedlmayr2012-kk} This self-returning process constitutes a Brownian bridge. Its root-mean square height scales as $h_\text{rms}=\sqrt{N/12}\Delta_\perp=\left[La_0\varepsilon/(3\sqrt{6})\right]^{1/2}$. (The factor $1/12$ for the bridge is derived, for example, in the Supplementary Material of Ref.~\onlinecite{Savio2016-dp}.) Our samples have $L=100$~nm. This gives $h_\text{rms}(\varepsilon)/\varepsilon^{1/2}=a_h$ with $a_h\approx 2.3$~nm in both cases. This values is of the same order of magnitude as our measured $a_h=6$~nm for NiCoFeTi and $a_h=4$~nm for Au. The other limit B of how steps are created on the surface is fully correlated. Each slip event occurs on the same slip plane and we end up with exactly two steps on the surface, up and down steps of the same height. The amplitude of the displacements then scales $h_\text{rms}\propto N$ rather than $h_\text{rms}\propto N^{1/2}$.

A generalization of process A is the sum of $N$ realizations of a random surface with a given Hurst exponent $H$. This leads to a progression of surfaces with $h_\text{rms}\propto N^{1/2}$, independent of $H$. Process B is the sum of $N$ identical realizations of a random surface which would trivially lead to $h_\text{rms}\propto N$.

We observe for the surface $h_\text{rms}\propto\varepsilon^{1/2}$ (process A) and for the bulk $u_\text{rms}\propto \varepsilon$ (process B). This appears to indicate that the surface and bulk processes discussed here are two limiting scenarios where the surface behaves close to the former and the bulk close to the latter. We believe the reason that the surface behaves differently lies in the fact that it moves perpendicular to the surface normal as the system is deformed. A progression of dislocations nucleating on identical slip systems and positions within the bulk therefore leave the surface at different locations. For CuZr, the process of accommodating deformation is different and this is manifested in smaller values for $a_h$ and $a_z$, yet the same scaling with $\varepsilon$. Further work is necessary to quantify the exact nature of the processes described here and derive a model for amorphous materials.

\begin{figure}
\begin{center}
\includegraphics[width=15cm]{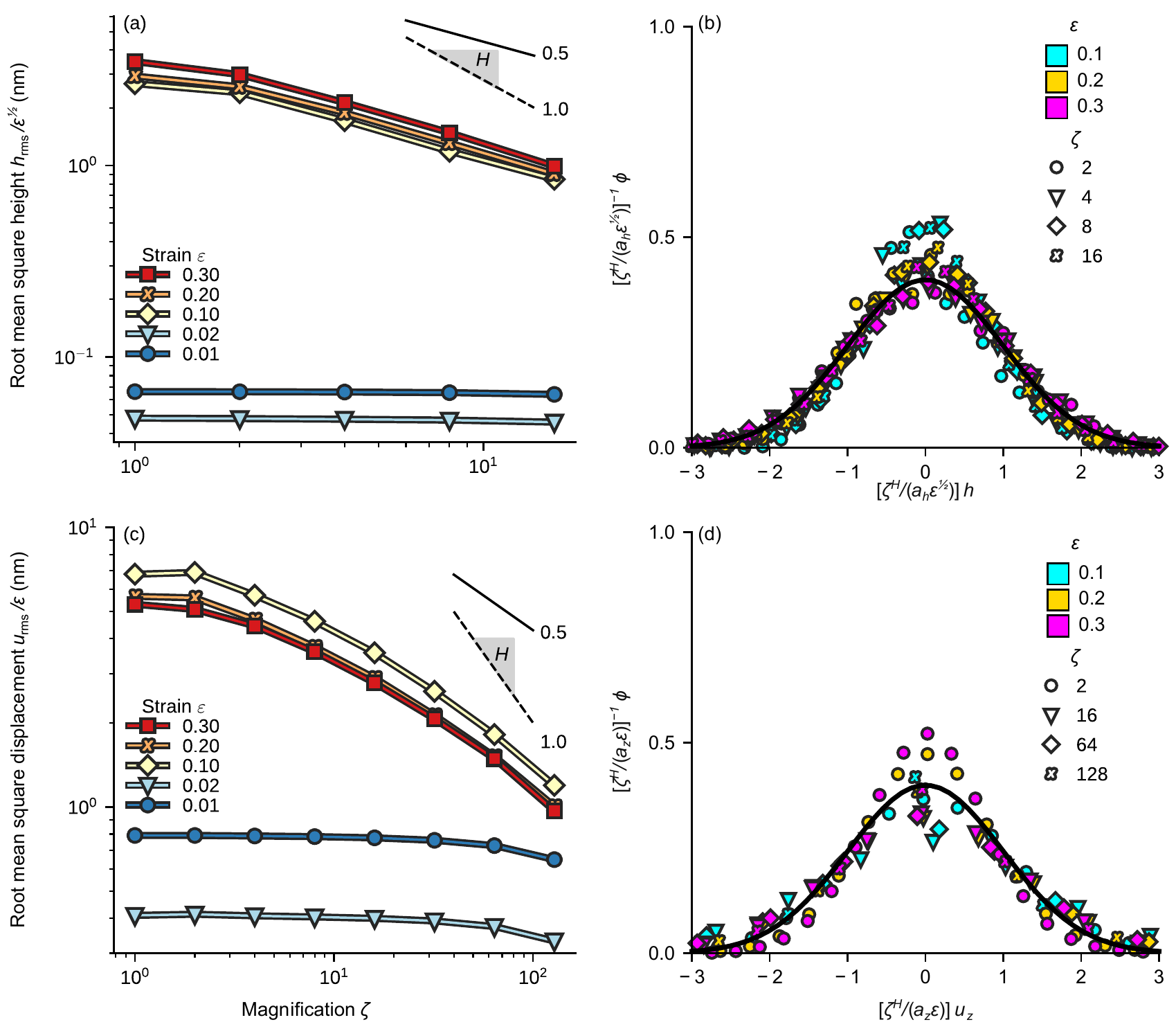}
\caption{Detailed analysis of the surface topography of Au. (a) Root-mean-square height $h_\text{rms}$ as a function of magnification $\zeta$ showing self-affine scaling over more than one decade in length. The data collapse in the plastic regime when normalized by $\varepsilon^{1/2}$, where epsilon is the strain due to compression. Panel (b) shows the underlying distribution function of heights $h$ at different $\varepsilon$, which collapses upon rescaling heights $h$ by $\zeta^H/(a_h \varepsilon^{1/2})$ and letting $a_h=4$~nm. (c) Root-mean-square amplitude $u_\text{rms}$ of the $z$-component of the subsurface displacement field $u_z$ as a function of $\zeta$ within the bulk. The displacement data collapses when normalized by $\varepsilon$. The bulk displacement field shows self-affine scaling over more than two decades in magnification. Panel (d) shows the underlying distribution function of the displacements $u_z$, which collapses upon rescaling displacements $u_z$ by $\zeta^H/(a_z \varepsilon)$ and letting $a_z=9$~nm.  Solid and dashed lines in (a) and (c) show perfect self-affine scaling for reference with $H=0.5$ and $H=1.0$, respectively. The solid lines in panels (b) and (d) show the standard normal distribution. The respective scaling collapses use the same Hurst exponent $H=0.5$.}
\end{center}
\end{figure}

\newpage

\begin{figure}
\begin{center}
\includegraphics[width=15cm]{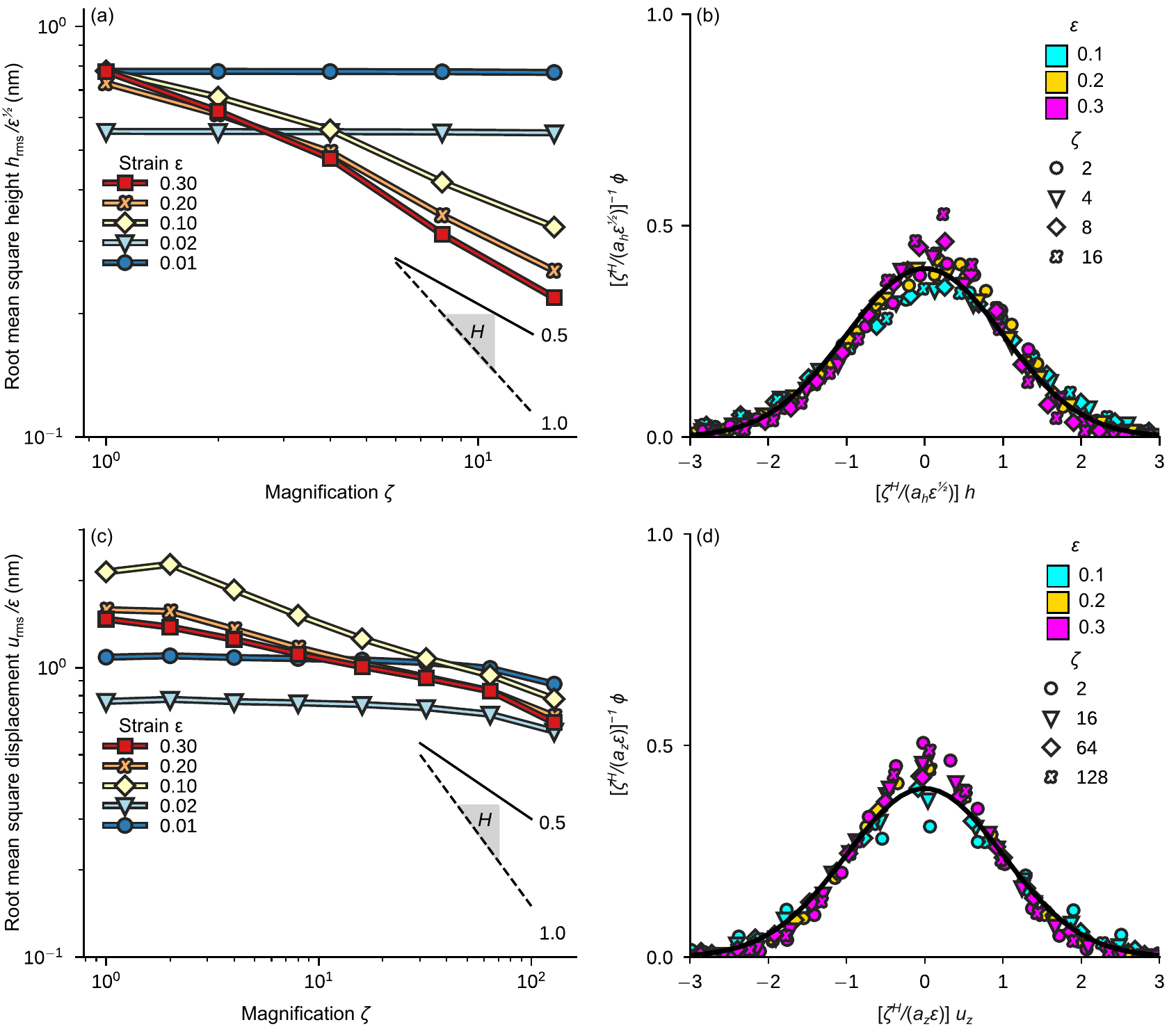}
\caption{Detailed analysis of the surface topography of CuZr. (a) Root-mean-square height $h_\text{rms}$ as a function of magnification $\zeta$ showing self-affine scaling over more than one decade in length. The data collapse in the plastic regime when normalized by $\varepsilon^{1/2}$, where epsilon is the strain due to compression. Panel (b) shows the underlying distribution function of heights $h$ at different $\varepsilon$, which collapses upon rescaling heights $h$ by $\zeta^H/(a_h \varepsilon^{1/2})$ and letting $a_h=0.8$~nm. (c) Root-mean-square amplitude $u_\text{rms}$ of the $z$-component of the subsurface displacement field $u_z$ as a function of $\zeta$ within the bulk. The displacement data collapse when normalized by $\varepsilon$. The bulk displacement field shows self-affine scaling over more than two decades in magnification. Panel (d) shows the underlying distribution function of the displacements $u_z$, which collapses upon rescaling displacements $u_z$ by $\zeta^H/(a_z \varepsilon)$ and letting $a_z=2$~nm.  Solid and dashed lines in (a) and (c) show perfect self-affine scaling for reference with $H=0.5$ and $H=1.0$, respectively. The solid lines in panels (b) and (d) show the standard normal distribution. Surface and bulk distribution functions do not collapse with the same Hurst exponent. Data shown here uses $H=0.4$ for the surface (panel b) and $H=0.2$ for the bulk (panel d).}
\end{center}
\end{figure}

\newpage

\begin{figure}
\begin{center}
\includegraphics[width=8cm]{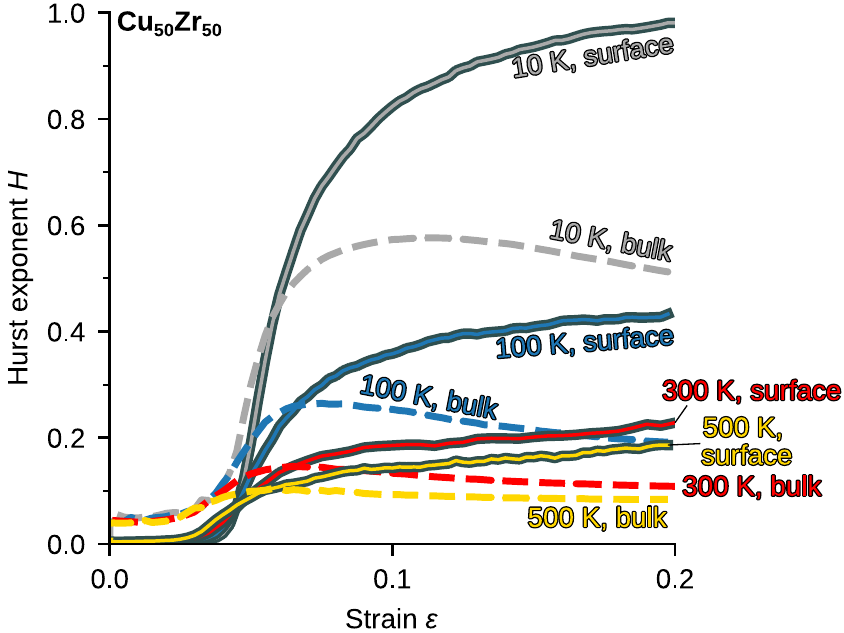}
\caption{Temperature dependence of the Hurst exponent for CuZr. The figure shows the evolution of the Hurst exponent of the surface (solid lines) and bulk (dashed lines) at the temperatures indicated as a function of applied strain $\varepsilon$. As the temperature approaches the glass transition temperature (around $800$~K), the surface roughness and bulk deformation becomes uncorrelated as indicated by a vanishing Hurst exponent.}
\end{center}
\end{figure}

\end{document}